\documentclass[11pt]{article}
\usepackage{amsfonts}
\usepackage{amsgen,amsmath,amstext,amsbsy,amsopn,amssymb}
\usepackage[pdftex]{graphicx}

\usepackage[usenames]{color}

\newtheorem{Definition}{Definition}[section]
\newtheorem{Theorem}{Theorem}[section]

\newtheorem{Lemma}{Lemma}[section]
\newtheorem{Remark}{Remark}[section]

\newtheorem{Proposition}{Proposition}[section]

\newcommand{\be}{\begin{equation}}
\newcommand{\ee}{\end{equation}}

\newcommand{\argmin}{\mathop{\rm arg\min}}

\newcommand{\cB}{\mathcal{B}}

\newcommand{\cM}{\mathcal{M}}

\newcommand{\RR}{\mathbb{R}}

\allowdisplaybreaks

\begin{document}

\title{Sparse Representation of a Polytope and Recovery of Sparse Signals and Low-rank Matrices\footnote{The research was supported in part by NSF FRG Grant DMS-0854973 and NIH Grant R01 CA127334-05.}}
\author{T. Tony Cai \, and \, Anru Zhang\\
Department of Statistics\\
The Wharton School\\
University of Pennsylvania}
\date{}
\maketitle

\begin{abstract}
This paper considers compressed sensing and affine rank minimization in both noiseless and noisy cases and establishes sharp restricted isometry conditions for sparse signal and low-rank matrix recovery.  The analysis relies on a key technical tool which represents points in a polytope by convex combinations of sparse vectors. The technique is elementary while leads to sharp results.

It is shown that for any given constant $t\ge {4/3}$, in compressed sensing $\delta_{tk}^A < \sqrt{(t-1)/t}$ guarantees the exact recovery of all $k$ sparse signals in the noiseless case through the constrained $\ell_1$ minimization, and similarly in affine rank minimization $\delta_{tr}^\mathcal{M}< \sqrt{(t-1)/t}$ ensures the exact reconstruction of  all matrices with rank at most $r$ in the noiseless case via the constrained nuclear norm minimization. Moreover, for any $\epsilon>0$, $\delta_{tk}^A<\sqrt{\frac{t-1}{t}}+\epsilon$ is not sufficient to guarantee the exact recovery of all $k$-sparse signals for large $k$. Similar result also holds for matrix recovery.  In addition, the conditions $\delta_{tk}^A < \sqrt{(t-1)/t}$  and $\delta_{tr}^\mathcal{M}< \sqrt{(t-1)/t}$ are also shown to be sufficient respectively for stable recovery of approximately sparse signals and low-rank matrices in the noisy case. 
\end{abstract}


\section{Introduction}

Efficient recovery of sparse signals and low-rank matrices has been a very active area of recent research in applied mathematics, statistics, and machine learning, with many important applications, ranging from signal processing \cite{Tropp, Davenport} to medical imaging \cite{Lustig} to radar systems \cite{Baraniuk_radar, Herman_radar}. A central goal is to develop fast algorithms that can recover sparse signals and low-rank matrices from a relatively small number of linear measurements. Constrained $\ell_1$-norm minimization and nuclear norm minimization are among the most well-known algorithms for the recovery of sparse signals and low-rank matrices respectively.

In compressed sensing, one observes
\begin{equation}\label{eq:modelsignal}
y=A\beta+z,
\end{equation}
where $y\in \mathbb{R}^n$,  $A\in \mathbb{R}^{n\times p}$ with $n\ll p$, $\beta\in \mathbb{R}^p$ is an unknown sparse signal, and $z\in\mathbb{R}^n$ is a vector of measurement errors. The goal is to recover the unknown signal $\beta\in\mathbb{R}^p$ based on the measurement matrix  $A$ and the observed signal $y$.
The constrained $\ell_1$ minimization method proposed by Cand\'es and Tao \cite{Candes_Decoding} estimates the signal $\beta$ by
\begin{equation}\label{eq:signalmini}
\hat\beta=\argmin_{\beta\in \RR^p} \{ \|\beta\|_1: \; \mbox{ subject to } \; A\beta-y\in\mathcal{B}\},
\end{equation}
where $\mathcal{B}$ is a set determined by the noise structure. In particular, $\cB$ is taken to be $\{0\}$ in the noiseless case. This constrained $\ell_1$ minimization method has now been well studied and it is understood that the procedure provides an efficient method for sparse signal recovery.

A closely related problem to compressed sensing is  the affine rank minimization problem (ARMP) (Recht et al. \cite{Recht_Matrix}), which aims to recover an unknown low-rank matrix based on its affine transformation. In ARMP, one observes
\begin{equation}\label{eq:modelmatrix}
b = \mathcal{M}(X) + z,
\end{equation}
where $\mathcal{M}:\mathbb{R}^{m\times n}\to \mathbb{R}^q$ is a known linear map, $X\in\mathbb{R}^{m\times n}$ is an unknown low-rank matrix of interest, and $z\in \mathbb{R}^q$ is measurement error. The goal is to recover the low-rank matrix $X$ based on the linear map $\cM$ and the observation $b\in \RR^q$.
Constrained nuclear norm minimization \cite{Recht_Matrix}, which is analogous to $\ell_1$ minimization in compressed sensing,  estimates $X$ by
\begin{equation}\label{eq:matrixmini}
 X_\ast=\argmin_{B\in \mathbb{R}^{m\times n}} \{\|B\|_\ast: \; \mbox{ subject to }\;  \mathcal{M}(B)-b\in\mathcal{B}\},
\end{equation}
where $\|B\|_\ast$ is the nuclear norm of $B$, which is defined as the sum of all singular values of $B$.  

One of the most widely used frameworks in compressed sensing is the  \emph{restrict isometry property} (RIP) introduced in Cand\'es and Tao \cite{Candes_Decoding}. A vector $\beta\in \RR^p$ is called $s$-sparse if $|$supp$(\beta)|\leq s$, where supp$(\beta) = \{i: \beta_i\neq0\}$ is the support of $\beta$.
\begin{Definition}\label{df:signalRIP}
Suppose $A\in\mathbb{R}^{n\times p}$ is a measurement matrix and $1\leq s\leq p$ is an integer. The restricted isometry constant (RIC) of order $s$ is defined as the smallest number $\delta_k^A$ such that for all $s$-sparse vectors $\beta\in \RR^p$, 
\begin{equation}
(1-\delta_s^A)\|\beta\|_2^2\leq\|A\beta\|_2^2\leq(1+\delta_s^A)\|\beta\|_2^2.
\end{equation}
When $s$ is not an integer, we define $\delta_s^A$ as $\delta_{\lceil s\rceil}^A$.
\end{Definition}

Different conditions on the RIC for sparse signal recovery have been introduced and studied in the literature. For example,  sufficient conditions for the exact recovery in the noiseless case include $\delta_{2k}<\sqrt{2}-1$ in \cite{Candes08}, $\delta_{2k}<0.472$ in \cite{Cai_Shift}, $\delta_{2k}<0.497$ in \cite{Mo}, $\delta_{k}<0.307$ in \cite{Cai_Newbound}, $\delta_{k}<1/3$ and $\delta_{2k}\leq 1/2$ in \cite{Cai_Zhang}. There are also other sufficient conditions that involve the RIC of different orders, e.g. $\delta_{3k}^A+3\delta_{4k}^A<2$ in \cite{Candes_incompletemeasurements}, $\delta_k^A+\delta_{2k}^A<1$ in \cite{Cai_Zhang2}, $\delta_{2k}^A<0.5746$ jointly with $\delta_{8k}^A<1$, $\delta_{3k}^A<0.7731$ jointly with $\delta_{16k}^A<1$ in \cite{Zhou} and $\delta_{2k}^A < 4/\sqrt{41}$ in \cite{Andersson}. 

Similar to the RIP for  the measurement matrix $A$ in compressed sensing  given in Definition \ref{df:signalRIP},  a restricted isometry property for a linear map $\cM$  in ARMP can be given. For two matrices $X$ and $ Y$ in $\mathbb{R}^{m\times n}$,  define their inner product as $\langle X, Y\rangle = \sum_{i,j} X_{ij}Y_{ij}$ and the Frobenius norm as $\|X\|_F = \sqrt{\langle X, X\rangle} = \sqrt{\sum_{i,j}X_{ij}^2}$. 
\begin{Definition}\label{df:matrixRIP}
Suppose $\mathcal{M}:\mathbb{R}^{n\times m}\to\mathbb{R}^q$ is a linear map and $1\leq r\leq \min(m, n)$ is an integer. The restricted isometry constant (RIC) of order $r$ for $\mathcal{M}$ is defined as the smallest number $\delta_r^\mathcal{M}$ such that for all matrices $X$ with rank at most $r$, 
\begin{equation}
(1-\delta_r^\mathcal{M})\|X\|_F^2\leq\|\mathcal{M}(X)\|_2^2\leq(1+\delta_r^\mathcal{M})\|X\|_F^2.
\end{equation}
When $r$ is not an integer, we define $\delta_r^\mathcal{M}$ as $\delta_{\lceil r\rceil}^\mathcal{M}$.
\end{Definition}
As in compressed sensing, there are many sufficient conditions based on the RIC to guarantee the exact recovery of matrices of rank at most $r$ through the constrained nuclear norm minimization \eqref{eq:matrixmini}. These include $\delta_{4r}^\mathcal{M}<\sqrt{2}-1$ \cite{Candes_Oracle}, $\delta_{5r}^\mathcal{M}<0.607$, $\delta_{4r}^\mathcal{M}<0.558$, and $\delta_{3r}^\mathcal{M}<0.4721$ \cite{Mohan}, $\delta_{2r}^{\mathcal{M}}<0.4931$ \cite{Wang_NewRIC}, $\delta_{r}^\mathcal{M}<0.307$ \cite{Wang_NewRIC},  $\delta_r^\mathcal{M}<1/3$  \cite{Cai_Zhang}, and $\delta_{2r}^\mathcal{M}< 1/2$ \cite{Cai_Zhang}. 

Among these sufficient RIP conditions, $\delta_k^A<1/3$ and $\delta_r^\mathcal{M}<1/3$ have been verified  in \cite{Cai_Zhang} to be sharp for both sparse signal recovery and low-rank matrix recovery problems. Sharp conditions on the higher order RICs are however still unknown. As pointed out by Blanchard and Thompson \cite{Blanchard}, higher-order RIC conditions can be satisfied by a significantly larger set of Gaussian random matrices in some settings. It is therefore of both theoretical and practical interests to obtain sharp sufficient conditions on the high order RICs.

\begin{figure}[htbp]
\begin{center}
  \includegraphics[width=5in,height=3.5in]{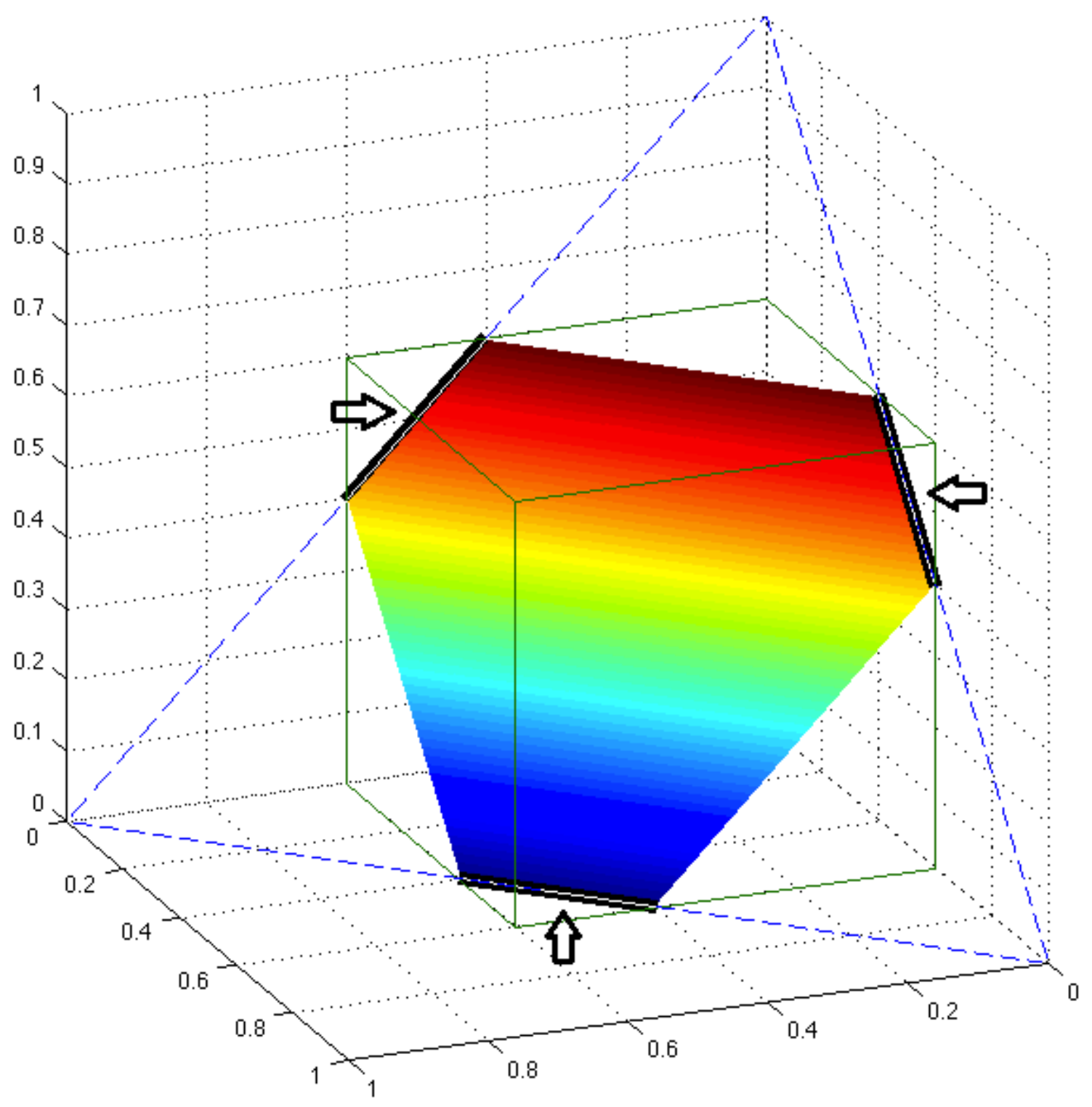}\\
  \caption{A graphical illustration of sparse representation of a polytope in one orthant with $p=3$ and $s=2$. All the points in the colored area can be expressed as convex combinations of the sparse vectors represented by the three pointed black line segments on the edges. }
\end{center}
\label{polytope.fig}
\end{figure}

In this paper, we develop a new elementary technique for the analysis of the constrained $\ell_1$-norm minimization and  nuclear norm minimization procedures and establish sharp RIP conditions on the high order RICs for sparse signal and low-rank matrix recovery. The analysis is surprisingly simple, while leads to sharp results. The key technical tool we develop states an elementary geometric fact: Any point in a polytope can be represented as a convex combination of sparse vectors.  The following lemma may be of independent interest.
\begin{Lemma}[Sparse Representation of a Polytope]
\label{lm:mean}
For a positive number $\alpha$ and a positive integer $s$, define the polytope $T(\alpha, s) \subset \RR^p$ by
\[
T(\alpha, s) =\{v\in \mathbb{R}^p: \|v\|_\infty\leq \alpha, \; \|v\|_1\leq s\alpha\}.
\]
For any $v\in \mathbb{R}^p$, define the set of sparse vectors $U(\alpha, s, v) \subset \RR^p$ by
\be
\label{eq:mean_u}
U(\alpha, s, v) =\{u\in \RR^p:\;  {\rm supp}(u) \subseteq {\rm supp}(v),\;  \|u\|_0\le s, \; \|u\|_1=\|v\|_1,\; \|u\|_\infty\leq \alpha\}.
\ee
Then $v \in T(\alpha, s)$ if and only if $v$ is in the convex hull of  $U(\alpha, s, v)$. In particular, any $v\in T(\alpha, s)$ can be expressed as 
$$v=\sum_{i=1}^N\lambda_i u_i, \quad\mbox{and }\; 0\leq \lambda_i\leq 1,\quad \sum_{i=1}^N \lambda_i=1, \quad\mbox{and } \; u_i \in U(\alpha, s, v).$$
\end{Lemma}
Lemma \ref{lm:mean} shows that  any  point $v\in\mathbb{R}^p$  with $\|v\|_\infty\leq \alpha$ and $\|v\|_1\leq s\alpha$ must lie in a convex polytope whose extremal points are $s$-sparse vectors $u$ with $\|u\|_1=\|v\|_1$ and $\|u\|_\infty \leq \alpha$, and vice versa. This geometric fact turns out to be a powerful tool in analyzing  constrained $\ell_1$-norm minimization for compressed sensing and nuclear norm minimization for ARMP,  since it represents a non-sparse vector by the sparse ones, which provides a bridge between general vectors and the RIP conditions. A graphical illustration of Lemma \ref{lm:mean} is given in Figure \ref{polytope.fig}.

\vspace{-10pt}

Combining the results developed in Sections  \ref{main.sec} and  \ref{matrix.sec}, we establish the following sharp sufficient RIP conditions for the exact recovery of all $k$-sparse signals and low-rank matrices in the noiseless case. We focus here on the exact sparse and noiseless case; the general approximately sparse (low-rank)  and noisy case is considered in Sections  \ref{main.sec} and  \ref{matrix.sec}.
\begin{Theorem}\label{th:main}
Let $y=A\beta$ where $\beta\in\mathbb{R}^p$ is a  $k$-sparse vector. If 
\be 
\label{eq:RIPcondition}
\delta_{tk}^A<\sqrt{\frac{t-1}{t}}
\ee
for some $t\ge 4/3$, then the $\ell_1$ norm minimizer $\hat\beta$ of \eqref{eq:signalmini} with $\mathcal{B}=\{0\}$ recovers $\beta$ exactly.

Similarly, suppose $b=\cM(X)$ where the matrix $X \in\mathbb{R}^{m\times n}$is of rank at most  $r$. If 
\be 
\label{eq:RIPcondition2}
\delta_{tr}^\cM<\sqrt{\frac{t-1}{t}}
\ee
for some $t\ge 4/3$,
then the nuclear norm minimizer $X_*$ of  \eqref{eq:matrixmini} with $\mathcal{B}=\{0\}$ recovers $X$ exactly.
\end{Theorem}

Moreover, it will be shown that for any $\epsilon>0$, $\delta_{tk}^A<\sqrt{\frac{t-1}{t}}+\epsilon$ is not sufficient to guarantee the exact recovery of all $k$-sparse signals for large $k$. Similar result also holds for matrix recovery. For the more general approximately sparse (low-rank) and noisy cases considered  in Sections  \ref{main.sec} and  \ref{matrix.sec}, it is shown that Conditions \eqref{eq:RIPcondition} and \eqref{eq:RIPcondition2} are also sufficient respectively  for  stable recovery of (approximately) $k$-sparse signals and (approximately) rank-$r$ matrices in the noisy case. An oracle inequality is also given in the case of compressed sensing with Gaussian noise under the condition $\delta_{tk}^A < \sqrt{(t-1)/t}$ when $t\geq 4/3$.

The rest of the paper is organized as follows.  Section \ref{main.sec} considers sparse signal recovery and Section \ref{matrix.sec} focuses on low-rank matrix recovery. Discussions on  the case  $t<4/3$ and some related issues are given in Section \ref{discussion.sec}. The proofs of the key technical result Lemma \ref{lm:mean} and the main theorems are contained in Section \ref{proof.sec}.

\section{Compressed Sensing}\label{main.sec}

We consider compressed sensing in this section and establish  the sufficient RIP condition $\delta_{tk}^A<\sqrt{(t-1)/t}$  in the noisy case which implies immediately the results in the noiseless case given in Theorem \ref{th:main}. For $v\in\mathbb{R}^p$, we denote $v_{\max(k)}$ as $v$ with all but the largest $k$ entries in absolute value set to zero, and $v_{-\max(k)}=v-v_{\max(k)}$. 

Let us consider the signal recovery model \eqref{eq:modelsignal} in the setting where the observations contain noise and the signal is not exactly $k$-sparse. This is  of significant  interest for many applications. Two types of bounded noise settings,
\[
z\in\mathcal{B}^{\ell_2}(\varepsilon) \triangleq \{z:\|z\|_2\leq \varepsilon\} \quad\mbox{and}\quad  z\in \mathcal{B}^{DS}(\varepsilon) \triangleq \{z:\|Az\|_\infty \leq \varepsilon\},
\]
are of particular interest.
The first bounded noise case was considered for example in \cite{DET}. The second case is motivated by the \emph{Dantzig Selector} procedure proposed in \cite{Candes_Dantzig}.
Results on the Gaussian noise case, which is commonly studied in statistics, follow immediately.
For notational convenience, we write $\delta$ for $\delta^A_{tk}$.
\begin{Theorem}\label{th:noisy}
Consider the signal recovery model \eqref{eq:modelsignal} with $\|z\|_2\leq \varepsilon$. Suppose $\hat\beta^{\ell_2}$ is the minimizer of \eqref{eq:signalmini} with $\mathcal{B}=\mathcal{B}^{\ell_2}(\eta)=\{z:\|z\|_2\leq\eta\}$ for some $\eta\geq \varepsilon$. If  $\delta=\delta_{tk}^A<\sqrt{(t-1)/t}$ for some $t\ge  4/3$, then
\begin{equation}\label{eq:hatbetal2}
\|\hat\beta^{\ell_2}-\beta\|_2\leq \frac{\sqrt{2(1+\delta)}}{1-\sqrt{t/(t-1)}\delta}(\varepsilon+\eta) + \left(\frac{\sqrt{2}\delta  + \sqrt{t(\sqrt{(t-1)/t}-\delta)\delta}}{t(\sqrt{(t-1)/t}-\delta)}+1\right)\frac{2\|\beta_{-\max(k)}\|_1}{\sqrt{k}}.
\end{equation} 
Now consider the signal recovery model \eqref{eq:modelsignal} with $\|A^Tz\|_\infty\leq \varepsilon$. Suppose $\hat\beta^{DS}$ is the minimizer of \eqref{eq:signalmini} with $\mathcal{B}=\mathcal{B}^{DS}(\eta)=\{z:\|A^Tz\|_\infty\leq\eta\}$ for some $\eta\geq \varepsilon$. If  $\delta=\delta_{tk}^A<\sqrt{(t-1)/t}$ for some $t\ge 4/3$, then
\begin{equation}\label{eq:hatbetaDS}
\|\hat\beta^{DS}-\beta\|_2\leq \frac{\sqrt{2tk}}{1-\sqrt{t/(t-1)}\delta}(\varepsilon+\eta) + \left(\frac{\sqrt{2}\delta + \sqrt{t(\sqrt{(t-1)/t}-\delta)\delta}}{t(\sqrt{(t-1)/t}-\delta)}+1\right)\frac{2\|\beta_{-\max(k)}\|_1}{\sqrt{k}}.
\end{equation} 
\end{Theorem}
\begin{Remark}{\rm
The result for the noiseless case follows directly from Theorem \ref{th:noisy}.
When $\beta$ is exactly $k$-sparse and there is no noise, by setting $\eta=\epsilon=0$ and by noting $\beta_{-\max(k)}=0$, we have $ \hat\beta= \beta$ from \eqref{eq:hatbetal2}, where $\hat \beta$ is the minimizer of \eqref{eq:signalmini} with $\mathcal{B}=\{0\}$.
}
\end{Remark}

\begin{Remark}
It should be noted that Theorems \ref{th:main} and \ref{th:noisy} also hold for $1<t < 4/3$ with exactly the same proof. However the bound $\sqrt{(t-1)/t}$  is not sharp for $1<t<4/3$. See Section \ref{discussion.sec} for further discussions. 
The condition $t\geq 4/3$ is crucial for the ``sharpness" results given in Theorem \ref{th:counterexample} at the end of this section. 
\end{Remark}

The signal recovery model \eqref{eq:modelsignal} with Gaussian noise is of particular interest in statistics and signal processing.   The following results on  the i.i.d. Gaussian noise case are immediate consequences of the above results on the bounded noise cases using the same argument as that in \cite{Cai_l1, Cai_Shift}, since the Gaussian random variables are essentially bounded.

\begin{Proposition}\label{pr:gaussian}
Suppose the error vector $z\sim N_n(0,\sigma^2I)$ in \eqref{eq:modelsignal}. $\delta_{tk}^A < \sqrt{(t-1)/t}$ for some $t\geq 4/3$. Let $\hat\beta^{\ell_2}$ be the minimizer of \eqref{eq:signalmini} with $\mathcal{B}=\{z:\|z\|_2\leq  \sigma\sqrt{n+2\sqrt{n\log n}}\}$ and let $\hat\beta^{DS}$ be the minimizer of \eqref{eq:signalmini} with $\mathcal{B} = \{z:\|A^Tz\|_\infty \leq  2\sigma\sqrt{\log p}\}$. Then with probability at least  $1-1/n$,
\begin{equation*}
\begin{split}
\|\beta^{\ell_2} - \beta\|_2\leq & \frac{2\sqrt{2(1+\delta)}}{1-\sqrt{t/(t-1)}\delta}\sigma\sqrt{n + 2\sqrt{n\log n}}\\
 & + \left(\frac{\sqrt{2}\delta  + \sqrt{t(\sqrt{(t-1)/t}-\delta)\delta}}{t(\sqrt{(t-1)/t}-\delta)}+1\right)\frac{2\|\beta_{-\max(k)}\|_1}{\sqrt{k}},
\end{split}
\end{equation*}
and with probability at least $1-1/\sqrt{\pi\log p}$,
\begin{equation*}
\begin{split}
\|\hat\beta^{DS}-\beta\|_2\leq &\frac{4\sqrt{2t}}{1-\sqrt{t/(t-1)}\delta}\sigma\sqrt{k \log p} \\
& + \left(\frac{\sqrt{2}\delta + \sqrt{t(\sqrt{(t-1)/t}-\delta)\delta}}{t(\sqrt{(t-1)/t}-\delta)}+1\right)\frac{2\|\beta_{-\max(k)}\|_1}{\sqrt{k}}.
\end{split}
\end{equation*}
\end{Proposition}

The oracle inequality approach was introduced by Donoho and Johnstone \cite{DJ} in the context of  wavelet thresholding for signal denoising. It provides an effective way to study the performance of an estimation procedure by comparing it to that of an ideal estimator. In the context of compressed sensing, oracle inequalities have been given in \cite{Cai_stable, Cai_Zhang, Candes_Dantzig, Candes_Oracle} under various settings. Proposition \ref{pr:oracle_signal} below provides an oracle inequality for compressed sensing with Gaussian noise under the condition $\delta_{tk}^A < \sqrt{(t-1)/t}$ when $t\geq 4/3$.
\begin{Proposition}\label{pr:oracle_signal}
Given \eqref{eq:modelsignal}, suppose the error vector $z\sim N_n(0, \sigma^2I)$, $\beta$ is $k$-sparse. Let $\hat \beta^{DS}$ be the minimizer of \eqref{eq:signalmini} with $\mathcal{B} = \{z:\|A^Tz\|_\infty \leq 4\sigma\sqrt{\log p}\}$. If $\delta_{tk}^A<\sqrt{(t-1)/t}$ for some $t\geq 4/3$, then with probability at least $1 - 1/\sqrt{\pi \log p}$,
\begin{equation}\label{eq:oracle_inequality}
\|\hat\beta^{DS} - \beta\|_2^2 \leq \frac{256t}{(1-\sqrt{t/(t-1)}\delta_{tk}^A)^2} \log p \sum_i \min(\beta_i^2, \sigma^2).
\end{equation}
\end{Proposition}

We now turn to show the sharpness of the condition $\delta_{tk}^A<\sqrt{(t-1)/t}$  for the exact recovery in the noiseless case and stable recovery in the noisy case. It should be noted tha tthe result in the special case $t=2$ was shown in \cite{Davies_RIPfail}.
\begin{Theorem}\label{th:counterexample}
Let $t\geq 4/3$. For all $\varepsilon>0$ and $k\geq 5/\varepsilon$, there exists a matrix $A$ satisfying $\delta_{tk}<\sqrt{\frac{t-1}{t}} +\varepsilon$ and some $k$-sparse vector $\beta_0$ such that
\begin{itemize}
\item in the noiseless case, i.e. $y=A\beta_0$, the $\ell_1$ minimization method \eqref{eq:signalmini} with $\mathcal{B}=\{0\}$ fail to exactly recover the $k$-sparse vector $\beta_0$, i.e. $\hat\beta\neq\beta_0$, where $\hat\beta$ is the solution to \eqref{eq:signalmini}.
\item in the noisy case, i.e. $y=A\beta_0+z$, for all constraints $\mathcal{B}_z$ (may depends on $z$), the $\ell_1$ minimization method \eqref{eq:signalmini}  fails to stably recover the $k$-sparse vector $\beta_0$, i.e. $\hat\beta \nrightarrow \beta$ as $z\to 0$, where $\hat\beta$ is the solution to \eqref{eq:signalmini}.
\end{itemize}
\end{Theorem}

\section{Affine Rank Minimization}\label{matrix.sec}

We consider the affine  rank minimization problem \eqref{eq:modelmatrix} in this section. As mentioned in the introduction,  this problem is closely related to compressed sensing. The close connections between compressed sensing and ARMP have been studied in Oymak, et al. \cite{Oymak11}. 
We shall  present here the analogous results on affine rank minimization without detailed proofs. 

For a matrix $X\in \mathbb{R}^{m\times n}$ (without loss of generality, assume that $m\leq n$) with the singular value decomposition $X=\sum_{i=1}^m a_iu_iv_i^T$ where the singular values $a_i$ are in descending order, we define $X_{\max(r)}=\sum_{i=1}^ra_iu_iv_i^T$ and $X_{-\max(r)}=\sum_{i=r+1}^m a_i u_i v_i^T$. We should also note that the nuclear norm $\|\cdot\|_\ast$ of a matrix equals the sum of the singular values, and the spectral norm $\|\cdot\|$ of a matrix equals its largest singular value. Their roles are similar to those of $\ell_1$ norm and $\ell_\infty$ norm in the vector case, respectively. For a linear operator $\mathcal{M}:\mathbb{R}^{m\times n}\to \mathbb{R}^q$,  its dual operator is denoted by $\mathcal{M}^\ast:\mathbb{R}^q\to \mathbb{R}^{m\times n}$. 

Similarly as in compressed sensing, we first consider the matrix recovery model \eqref{eq:modelmatrix} in  the case where the error  vector $z$ is in bounded sets: $\|z\|_2\leq\epsilon$ and $\|\mathcal{M}^\ast(z)\|\leq \varepsilon$. The corresponding nuclear norm minimization methods are given by \eqref{eq:matrixmini} with $\mathcal{B}=\mathcal{B}^{\ell_2}(\eta)$ and $\mathcal{B}=\mathcal{B}^{DS}(\eta)$ respectively, where
\begin{eqnarray}
\mathcal{B}^{\ell_2}(\eta) & =& \{z:\|z\|_2\leq\eta\},\label{eq:B^l2}\\
\mathcal{B}^{DS}(\eta) &=&\{z:\|\mathcal{M}^*(z)\|\leq\eta\} \label{eq:B^DS}.
\end{eqnarray}

\begin{Proposition}\label{pr:noisy}
Consider ARMP \eqref{eq:modelmatrix} with $ \|z\|_2\leq\varepsilon$. Let $X_\ast^{\ell_2}$ be the minimizer of \eqref{eq:matrixmini} with $\mathcal{B}=\mathcal{B}^{\ell_2}(\eta)$ defined in \eqref{eq:B^l2} for some $\eta\ge \epsilon$. If $\delta_{r}^{\mathcal{M}}<\sqrt{(t-1)/t}$ with $t\geq 4/3$, then
\begin{equation}
\|X_\ast^{\ell_2}-X\|_F\leq\frac{\sqrt{2(1+\delta)}}{1-\sqrt{t/(t-1)}\delta}(\varepsilon+\eta) + \left(\frac{\sqrt{2}\delta  + \sqrt{t(\sqrt{(t-1)/t}-\delta)\delta}}{t(\sqrt{(t-1)/t}-\delta)}+1\right)\frac{2\|X_{-\max(r)}\|_1}{\sqrt{r}}.
\end{equation}
Similarly, consider ARMP \eqref{eq:modelmatrix} with  $z$ satisfying $\|\mathcal{M}^\ast (z)\| \leq \varepsilon$. Let $X_\ast^{DS}$ be the minimizer of \eqref{eq:matrixmini} with $\mathcal{M}=\mathcal{B}^{DS}(\eta)$ defined in \eqref{eq:B^DS}, then
\begin{equation}
\|X_\ast^{DS}-X\|_F\leq\frac{\sqrt{2tr}}{1-\sqrt{t/(t-1)}\delta}(\varepsilon+\eta) + \left(\frac{\sqrt{2}\delta  + \sqrt{t(\sqrt{(t-1)/t}-\delta)\delta}}{t(\sqrt{(t-1)/t}-\delta)}+1\right)\frac{2\|X_{-\max(r)}\|_1}{\sqrt{r}}.
\end{equation}
\end{Proposition}


In the special noiseless case where $z=0$, it can be seen from either of these two inequalities above that all matrices $X$ with rank at most $r$ can be exactly recovered provided that $\delta_{tr}^{\mathcal{M}}<\sqrt{(t-1)/t}$, for some $t\ge 4/3$.

The following result shows that the condition $\delta_{tr}^{\mathcal{M}}<\sqrt{(t-1)/t}$ with $t\geq 4/3$ is sharp. These results together establish the optimal bound on $\delta_{tr}^\mathcal{M}$ $(t\geq 4/3)$ for the exact recovery in the noiseless case.
\begin{Proposition}\label{pr:counterexample}
Suppose $t\geq 4/3$. For all $\varepsilon>0$ and $r \geq 5/\varepsilon$, there exists a linear map $\mathcal{M}$ with $\delta_{tr}^{\mathcal{M}}< \sqrt{(t-1)/t} + \varepsilon$ and some matrix $X_0$ of rank at most $r$ such that
\begin{itemize}
\item in the noiseless case, i.e. $b = \mathcal{M}(X_0)$, the nuclear norm minimization method \eqref{eq:matrixmini} with $\mathcal{B}=\{0\}$ fails to exactly recover $X_0$, i.e. $X_\ast\neq X_0$, where $X_\ast$ is the solution to \eqref{eq:matrixmini}.
\item in the noisy case, i.e. $b = \mathcal{M}(X_0)+z$, for all constraints $\mathcal{B}_z$ (may depends on $z$), the nuclear norm minimization method \eqref{eq:matrixmini} fails to stably recover $X_0$, i.e. $X_\ast\nrightarrow X_0$ as $z\to 0$, where $X_\ast$ is the solution to \eqref{eq:matrixmini} with $\mathcal{B} = \mathcal{B}_z$.
\end{itemize}
\end{Proposition}

\section{Discussion}
\label{discussion.sec}
 
We shall focus the discussions in this section exclusively on compressed sensing as the results on affine rank minimization is  analogous.
In Section \ref{main.sec}, we have established the sharp RIP condition on the high-order RICs,
$$\delta_{tk}^A<\sqrt{\frac{t-1}{t}}  \quad \mbox{for some $t\ge \frac{4}{3}$,}$$
for the recovery of $k$-sparse signals in compressed sensing. In addition, it is known from \cite{Cai_Zhang} that $\delta_k^A<1/3$ is also a sharp RIP condition. For a general $t>0$, denote the sharp bound for $\delta_{tk}^A$ as $\delta_\ast(t)$. Then
$$\delta_\ast(1) = 1/3 \quad\mbox{and}\quad \delta_\ast(t)=\sqrt{(t-1)/t},\quad t\geq 4/3.$$ 
A natural question is: What is the value of $\delta_\ast(t)$ for $t<4/3$ and $ t\neq 1$? That is, what is the sharp bound for $\delta_{tk}^A$ when $t<4/3$ and $t\neq 1$? We have the following partial answer to the question.
\begin{Proposition}\label{pr:t<1}
Let $y = A\beta$ where $\beta\in\mathbb{R}^p$ is $k$-sparse. Suppose $0<t<1$ and $tk\geq 0$ to be an integer
\begin{itemize}
\item When $tk$ is even and $\delta_{tk}^A<\frac{t}{4-t}$, the $\ell_1$ minimization \eqref{eq:signalmini} with $\mathcal{B}=\{0\}$ recovers $\beta$ exactly.
\item When $tk$ is odd and $\delta_{tk}^A<\frac{\sqrt{t^2 - 1/k^2}}{4-2t+\sqrt{t^2-1/k^2}}$, the $\ell_1$ minimization $\eqref{eq:signalmini}$ with $\mathcal{B}=\{0\}$ recovers $\beta$ exactly.
\end{itemize}
\end{Proposition}

In addition, the following result shows that $\delta_*(t) \le \frac{t}{4-t}$ for all $0<t<4/3$. In particular,  when $t=1$, the upper bound $t/(4-t)$ coincides with the true sharp bound $1/3$. 
\begin{Proposition}\label{pr:t<4/3}
For $0<t<4/3$, $\varepsilon>0$ and any integer $k\geq1$, $\delta_{tk}^A<\frac{t}{4-t}+\varepsilon$ is not suffient for the exact recovery. Specifically, there exists a matrix $A$ with $\delta_{tk}^A=\frac{t}{4-t}$ and a $k$-sparse vector $\beta_0$ such that $\hat\beta\neq \beta_0$, where $\hat\beta$ is the minimizer of \eqref{eq:signalmini} with $\mathcal{B}=\{0\}$.
\end{Proposition}
Propositions \ref{pr:t<1} and \ref{pr:t<4/3} together show that $\delta_*(t) = \frac{t}{4-t}$ when $tk$ is even and  $0<t < 1$.
We are not able to provide a complete answer for $\delta_*(t)$ when $0<t<4/3$. We conjecture that $\delta_*(t) = \frac{t}{4-t}$ for all $0<t<4/3$.  The following figure plots $\delta_*(t) $ as a function of $t$ based on this conjecture for the interval $(0, 4/3)$.
\begin{figure}[htbp]
\begin{center}
  \includegraphics[width=5in,height=2.5in]{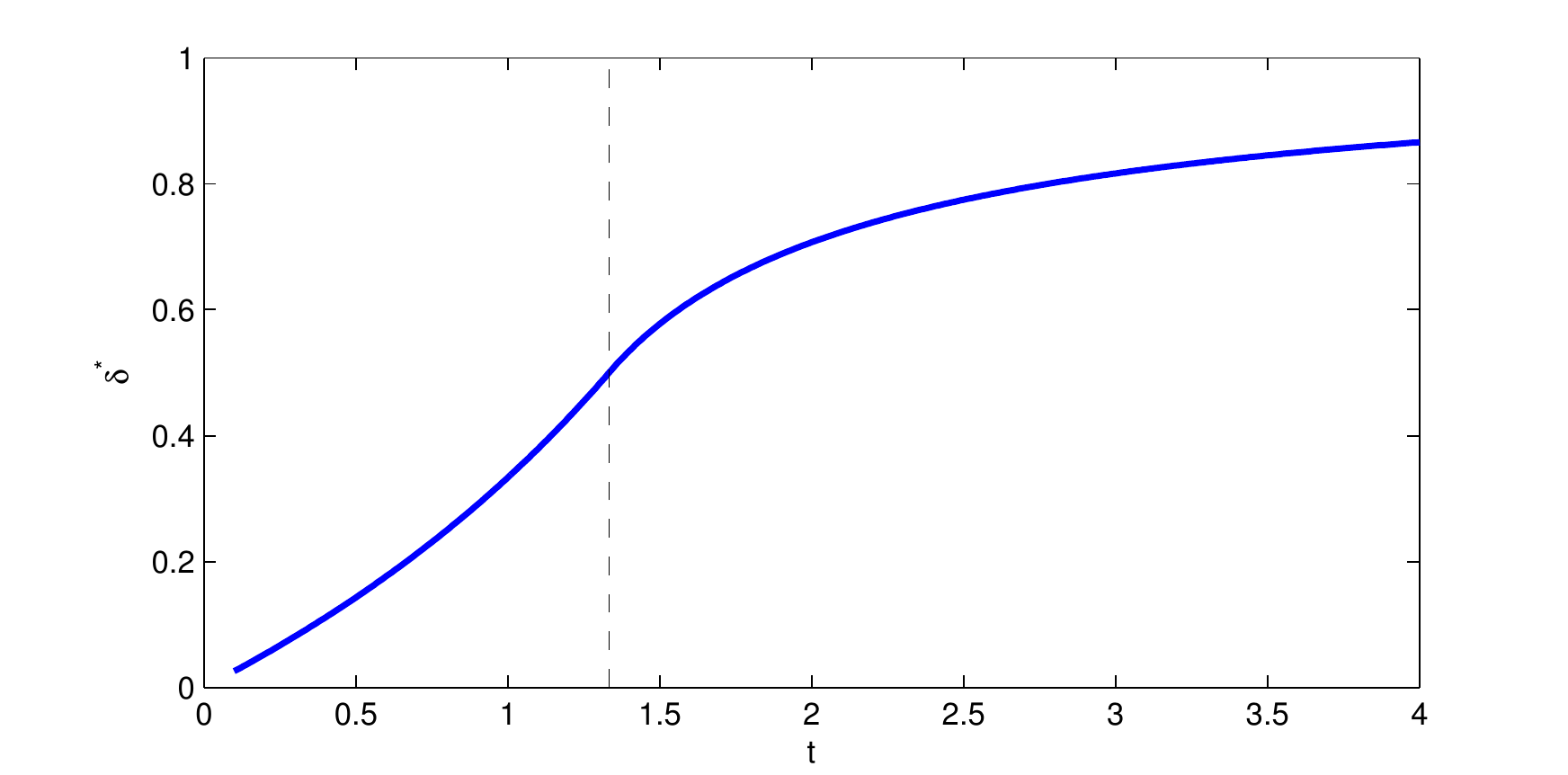}\\
  \caption{Plot of $\delta_\ast$ as a function of  $t$. The dotted line is $t=4/3$.}
\end{center}
\end{figure}

Our results show that exact recovery of $k$-sparse signals in the noiseless case is guaranteed if $\delta_{tk}^A<\sqrt{(t-1)/t}$ for some $t\ge 4/3$. It is then natural to ask the question: Among all these RIP conditions $\delta_{tk}^A<\delta_\ast(t)$,
which one is easiest to be satisfied? There is no general answer to this question as no condition is strictly weaker or stronger than the others. It is however interesting to consider special random measurement matrices $A=(A_{ij})_{n\times p}$ where
$$ A_{ij} \sim \mathcal{N}(0, 1/n), \quad A_{ij} \sim\left\{\begin{array}{ll}
1/\sqrt{n} & \text{w.p.} 1/2\\
-1/\sqrt{n} & \text{w.p.} 1/2
\end{array}\right. , \; \text{ or } \quad A_{ij} \sim \left\{\begin{array}{ll}
\sqrt{3/n} & \text{w.p.} 1/6\\
0 & \text{w.p.} 1/2\\
-\sqrt{3/n} & \text{w.p.} 1/6
\end{array}\right.
.$$ 
Baraniuk et al \cite{Baraniuk} provides a bound on RICs for a set of random matrices from concentration of measure.  For these random measurement matrices, Theorem 5.2 of \cite{Baraniuk} shows that
for positive integer $m<n$ and $0<\lambda<1$, 
\begin{equation}\label{eq:delta_probability}
 P(\delta_{m}^A < \lambda) \geq 1-2\left(\frac{12e p}{m \lambda}\right)^m\exp\left(-n(\lambda^2/16 - \lambda^3/48)\right).
\end{equation}
Hence,
for $t\geq4/3$,
$$ P(\delta_{tk}^A < \sqrt{(t-1)/t}) \geq 1-2\exp\left(tk\left(\log (12e/\sqrt{t(t-1)}) +\log(p/k)\right) - n\left(\frac{t-1}{16t} - \frac{(t-1)^{3/2}}{48t^{3/2}}\right)\right).$$
For $0<t<4/3$, using the conjectured value $\delta_*(t) = \frac{t}{4-t}$, we have
$$ P(\delta_{tk}^A < t/(4-t)) \geq 1 - 2\exp\left(tk(\log(12(4-t)e/t^2) + \log(p/k)) -n\left(\frac{t^2}{16(4-t)^2} - \frac{t^3}{48(4-t)^3}\right)\right).$$
It is easy to see when $p, k,$ and $p/k \to\infty$, the lower bound of $n$ to ensure $\delta_{tk}^A<t/(4-t)$ or $\delta_{tk}^A<\sqrt{(t-1)/t}$ to hold in high probability is $n\geq k\log(p/k) n^\ast(t)$, where
$$ n^\ast \triangleq \left\{\begin{array}{ll} t/\left(\frac{t^2}{16(4-t)^2}-\frac{t^3}{48(4-t)^3}\right) & t<4/3;\\
t/\left(\frac{t-1}{16t}-\frac{(t-1)^{3/2}}{48t^{3/2}}\right), & t\geq4/3.
\end{array}\right.$$
For the plot of $n^\ast(t)$, see Figure 1. $n^\ast(t)$ has minimum $83.2$ when $t = 1.85$. Moreover, among integer $t$, $t=2$ can also provide a near-optimal minimum: $n^\ast(2) = 83.7$. 

\begin{figure}[htbp]
\begin{center}
  \includegraphics[width=5in,height=2.5in]{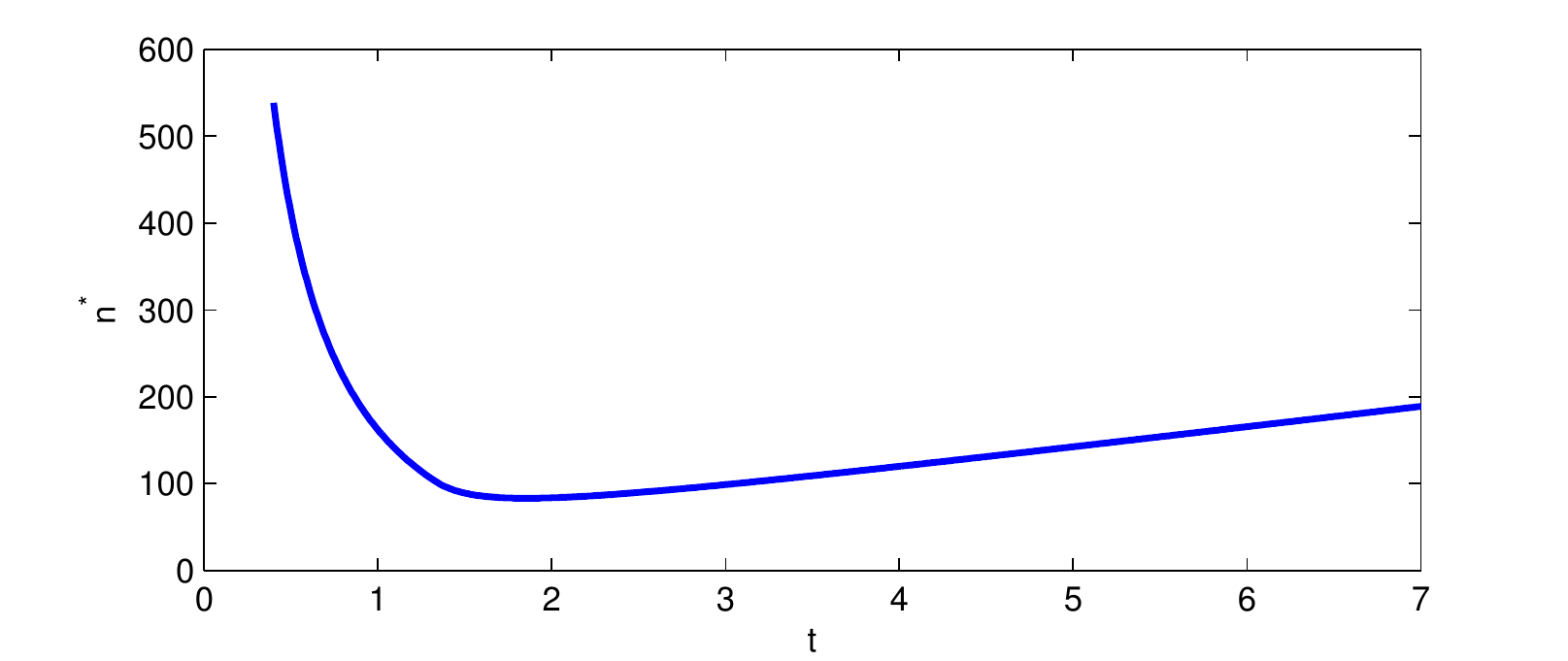}\\
  \caption{Plot of $n_\ast$ as a function of  $t$.}
\end{center}
\end{figure}
  
We should note that the above analysis is based on the bound given in \eqref{eq:delta_probability} which  itself can be possibly improved.

\section{Proofs}
\label{proof.sec}

We shall first establish the technical result, Lemma \ref{lm:mean}, and then prove the main results.

\medskip\noindent
{\bf Proof of Lemma \ref{lm:mean}.} 
First, suppose $v\in T(\alpha, s)$. We can prove $v$ is in the convex hull of $U(\alpha, s, v)$ by induction. If $v$ is $s$-sparse, $v$ itself is in $U(\alpha, s, v)$.

Suppose the statement is true for all $(l-1)$-sparse vectors $v$ ($l-1\geq s$). Then for any $l$-sparse vector $v$ such that $\|v\|_\infty\leq \alpha$, $\|v\|_1\leq s\alpha$, without loss of generality we assume that $v$ is not $(l-1)$-sparse (otherwise the result holds by assumption of $l-1$). Hence we can express $v$ as $v=\sum_{i=1}^l a_i e_i$, where $e_i$'s are different unit vectors with one entry of $\pm 1$ and other entries of zeros; $a_1\geq a_2\geq\cdots\geq a_l> 0$. Since $\sum_{i=1}^l a_i =\|v\|_1\leq s\alpha$, so
$$1\in D\triangleq\{1\leq j\leq l-1:a_j+a_{j+1}+\cdots+a_l\leq (l-j)\alpha\},$$
which means $D$ is not empty. Take the largest element in $D$ as $j$, which implies
 \begin{equation}\label{eq:thetaineq1}
 a_j+a_{j+1}+\cdots+a_l\leq (l-j)\alpha,\quad a_{j+1}+a_{j+2}+\cdots+a_l>(l-j-1)\alpha.
 \end{equation}
(It is noteworthy that even if the largest $j$ in $D$ is $l-1$, \eqref{eq:thetaineq1} still holds). Define
 \begin{equation}
 b_w\triangleq\frac{\sum_{i=j}^l a_i}{l-j}-a_w,\quad j\leq w\leq l,
 \end{equation}
which satisfies $\sum_{i=j}^la_i=(l-j)\sum_{i=j}^lb_i$. By \eqref{eq:thetaineq1}, for all $j\leq w\leq l$,
$$b_w\geq b_j=\frac{\sum_{i=j+1}^la_i}{l-j}-\frac{l-j-1}{l-j}a_j\geq\frac{\sum_{i=j+1}^la_i-(l-j-1)\alpha}{l-j}>0.$$
In addition, we define
\begin{equation}
v_w\triangleq\sum_{i=1}^{j-1}a_ie_i+(\sum_{i=j}^l b_i)\sum_{i=j,i\neq w}^l e_i\in R^{p},\quad \lambda_w \triangleq \frac{b_w}{\sum_{i=j}^l b_i},\quad j\leq w\leq l,
\end{equation}
then $0\leq\lambda_w\leq 1$, $\sum_{w=j}^l \lambda_w = 1$, $\sum_{w=j}^l\lambda_w v_w=v$, $\text{supp}(v_w)\subseteq\text{supp}(v)$. We also have
$$\|v_w\|_1=\sum_{i=1}^{j-1}a_i+(l-j)\sum_{w=j}^lb_w=\sum_{i=1}^{j-1}a_i+\sum_{i=j}^la_i=\|v\|_1,$$
$$\|v_w\|_\infty=\max\{a_1,\cdots, a_{j-1},\sum_{i=j}^l b_i\}\leq \max\{\alpha,\frac{\sum_{i=j}^l a_i}{l-j}\}\leq\alpha.$$
The last inequality is due to the first part of \eqref{eq:thetaineq1}. Finally, note that $v_w$ is $(l-1)$-sparse, we can use the induction assumption to find $\{u_{i,w}\in \mathbb{R}^p, \lambda_{i,w}\in \mathbb{R}: 1\leq i\leq N_w, j \leq w\leq l\}$ such that 
$$ u_{i,w} \text{ is } s\text{-sparse}, \quad \text{supp}(u_{i,w})\subseteq\text{supp}(v_i)\subseteq\text{supp}(v),\quad \|u_{i,w}\|_1=\|v_i\|_1=\|v\|_1,\quad \|u_{i,w}\|_\infty\leq \alpha;$$ 
In addition, $v_i = \sum_{i=1}^{N_w}\lambda_{i,w}u_{i,w}$, so $v = \sum_{w=j}^{l}\sum_{i=1}^{N_w} \lambda_w\lambda_{i,w} u_{i,w}$, which proves the result for $l$.

The proof of the other part of the lemma is easier. When $v$ is in the convex hull of $U(\alpha, s, v)$, then we have
$$\|v\|_\infty  = \|\sum_{i=1}^N \lambda_i u_i\|_\infty \leq \sum_{i=1}^N\lambda_i \|u_i\|_\infty\leq \alpha,$$
$$\|v\|_1 = \|\sum_{i=1}^N \lambda_i u_i\|_1 \leq \sum_{i=1}^N\lambda_i\|u_i\|_1 \leq \sum_{i=1}^N\lambda_i \|u_i\|_0\|u_i\|_\infty \leq s\alpha,$$
which finished the proof of the lemma. \quad $\square$

\medskip\noindent
{\bf Proof of Theorem \ref{th:main}} First, we assume that $tk$ is an integer.
By the well-known Null Space Property (Theorem 1 in \cite{Stojnic}), we only need to check for all $h\in\mathcal{N}(A)\setminus\{0\}$, $\|h_{\max(k)}\|_1<\|h_{-\max(k)}\|_1$.
Suppose there exists $h\in\mathcal{N}(A)\setminus\{0\}$, such that $\|h_{\max(k)}\|_1\geq \|h_{-\max(k)}\|_1$.  Set $\alpha=\|h_{\max(k)}\|_1/k$. We divide $h_{-\max(k)}$ into two parts, $h_{-\max(k)}=h^{(1)}+h^{(2)}$, where
$$h^{(1)}=h_{-\max(k)}\cdot 1_{\{i||h_{-\max(k)}(i)|> \alpha/(t-1)\}},\quad h^{(2)}=h_{-\max(k)}\cdot 1_{\{i||h_{-\max(k)}(i)|\leq \alpha/(t-1)\}}.$$
Then $\|h^{(1)}\|_1\leq \|h_{-\max(k)}\|_1\leq \alpha k$. Denote $|\text{supp}(h^{(1)})|= \|h^{(1)}\|_0=m$. Since all non-zero entries of $h^{(1)}$  have magnitude larger than $\alpha/(t-1)$, we have
$$\alpha k \geq \|h^{(1)}\|_1 = \sum_{i\in \text{supp}(h^{(1)})} |h^{(1)}(i)| \geq \sum_{i\in \text{supp}(h^{(1)})} \alpha/(t-1) = m\alpha/(t-1). $$
Namely $m\leq k(t-1)$. In addition we have
\begin{equation}\label{eq:main1}
\begin{split}
\|h^{(2)}\|_1& =\|h_{-\max(k)}\|_1-\|h^{(1)}\|_1  \leq k\alpha - \frac{m\alpha}{t-1}=(k(t-1)-m)\cdot \frac{\alpha}{t-1}, \\
\|h^{(2)}\|_\infty & \leq \frac{\alpha}{t-1}.
\end{split}
\end{equation}
We now apply Lemma \ref{lm:mean} with  $s=k(t-1)-m$. Then $h^{(2)}$  can be expressed as a convex combination of sparse vectors: $h^{(2)}=\sum_{i=1}^N\lambda_i u_i$, where $u_i$ is $(k(t-1)-m)$-sparse and 
\begin{equation}\label{eq:after_lemma_u_i}
\|u_i\|_1 = \|h^{(2)}\|_1, \quad \|u_i\|_\infty \leq \frac{\alpha}{(t-1)},\quad  \text{supp}(u_i) \subseteq \text{supp}(h^{(2)}).   
\end{equation}
Hence,
\begin{equation}\label{eq:l2_u_i}
\|u_i\|_2\leq\sqrt{\|u_i\|_0}\|u_i\|_\infty \leq\sqrt{k(t-1)-m}\|u_i\|_\infty\leq\sqrt{k(t-1)}\|u_i\|_\infty \leq\sqrt{k/(t-1)}\alpha.
\end{equation}
Now we suppose $\mu\geq0, c\geq0$ are to be determined. Denote $\beta_i=h_{\max(k)}+h^{(1)}+\mu u_i$, then 
\begin{equation}\label{eq:mainth2}
\sum_{j=1}^N\lambda_j \beta_j-c\beta_i=h_{max{(k)}}+h^{(1)}+\mu h^{(2)}-c\beta_i = (1-\mu-c) (h_{\max(k)}+h^{(1)})-c\mu u_i+\mu h.
\end{equation}
Since $h_{\max(k)}$, $h^{(1)}$, $u_i$ are $k$-, $m$-, $(k(t-1)-m)$-sparse respectively, $\beta_i = h_{\max(k)} + h^{(1)} + \mu u_i$, $\sum_{j=1}^N\lambda_j \beta_j-c\beta_i-\mu h = (1-\mu-c)(h_{\max(k)}+h^{(1)}) - c\mu u_i$ are all $tk$-sparse vectors.

We can check the following identity in $\ell_2$ norm,
\begin{equation}\label{eq:thmain_identity}
\begin{split}
\sum_{i=1}^N\lambda_i\|A(\sum_{j=1}^N\lambda_j\beta_j -c\beta_i)\|_2^2 + (1-2c)\sum_{1\leq i<j\leq N}\lambda_i\lambda_j\|A(\beta_i-\beta_j)\|_2^2
=\sum_{i=1}^N \lambda_i(1-c)^2\|A\beta_i\|_2^2.
\end{split}
\end{equation}
Since $Ah=0$ and \eqref{eq:mainth2}, we have $A(\sum_{j=1}^N\lambda_j\beta_j - c\beta_i) = A((1-\mu-c)(h_{\max(k)}+h^{(1)})-c\mu u_i)$. Set $c=1/2$, $\mu=\sqrt{t(t-1)}-(t-1)$, let the left hand side of \eqref{eq:thmain_identity} minus the right hand side, we get
\begin{eqnarray*}
0 & \leq & (1+\delta_{tk}^A) \sum_{i=1}^N\lambda_i\left((1-\mu-c)^2\|h_{\max(k)}+h^{(1)}\|_2^2+c^2\mu^2\|u_i\|_2^2\right)\\
 &          & -(1-\delta_{tk}^A)\sum_{i=1}^N\lambda_i(1-c)^2\left(\|h_{\max(k)}+h^{(1)}\|_2^2+\mu^2\|u_i\|_2^2\right)\\
   & = & \sum_{i=1}^N\lambda_i\left[\left((1+\delta_{tk}^A)(\frac{1}{2}-\mu)^2-(1-\delta_{tk}^A)\cdot\frac{1}{4}\right)\|h_{\max(k)}+h^{(1)}\|_2^2+\frac{1}{2}\delta_{tk}^A\mu^2\|u_i\|_2^2\right]\\
   & \leq & \sum_{i=1}^N\lambda_i\|h_{\max(k)}+h^{(1)}\|_2^2\left[(\mu^2-\mu)+\delta_{tk}^A\left(\frac{1}{2}-\mu +(1+\frac{1}{2(t-1)})\mu^2\right)\right]\\
   & = & \|h_{\max(k)}+h^{(1)}\|_2^2 \left[\delta_{tk}^A\left((2t-1)t-2t\sqrt{t(t-1)}\right)-\left((2t-1)\sqrt{t(t-1)}-2t(t-1)\right)\right]\\
   & < & 0.
\end{eqnarray*}
We used the fact that 
$$\delta_{tk}^A<\sqrt{(t-1)/t},$$
$$\quad \|u_i\|_2\leq \sqrt{k/(t-1)}\alpha\leq \frac{\|h_{\max(k)}\|_2}{\sqrt{(t-1)}}\leq\frac{\|h_{\max(k)}+h^{(1)}\|_2}{\sqrt{t-1}}$$ above. This is a contradiction.

When $tk$ is not an integer, note $t' = \lceil tk \rceil/k$, then $t'>t$, $t'k$ is an integer,
$$\delta_{t'k}=\delta_{tk} < \sqrt{\frac{t-1}{t}} < \sqrt{\frac{t'k-1}{t'k}},$$
which can be deduced to the former case. Hence we finished the proof.\quad $\square$

\medskip\noindent
{\bf Proof of Theorem \ref{th:noisy}.}
We first prove the inequality on $\hat\beta^{\ell_2}$ \eqref{eq:hatbetal2}. Again, we assume that $tk$ is an integer at first.
Suppose $h=\hat\beta^{\ell_2}-\beta$, we shall use a widely known result (see, e.g., \cite{Cai_l1}, \cite{Candes_Dantzig}, \cite{Candes_incompletemeasurements}, \cite{Donoho}), 
$$\|h_{-\max(k)}\|_1\leq \|h_{\max(k)}\|_1 +2\|\beta_{-\max(k)}\|_1.$$
Besides,
\begin{equation}\label{eq:Ah}
\|Ah\|_2\leq \|y-A\beta\|_2+\|A\hat\beta^{\ell_2}-y\|_2 \leq \varepsilon+\eta.
\end{equation}
Define $\alpha=(\|h_{\max(k)}\|_1+2\|\beta_{-\max(k)}\|_1)/k$. Similarly as the proof of Theorem \ref{th:main}, we divide $h_{-\max(k)}$ into two parts, $h_{-\max(k)}=h^{(1)}+h^{(2)}$, where
$$h^{(1)}=h_{-\max(k)}\cdot 1_{\{i||h_{-\max(k)}(i)|> \alpha/(t-1)\}},\quad h^{(2)}=h_{-\max(k)}\cdot 1_{\{i||h_{-\max(k)}(i)|\leq \alpha/(t-1)\}}.$$
Then $\|h^{(1)}\|_1\leq \|h_{-\max(k)}\|_1\leq \alpha k$. Denote $|\text{supp}(h^{(1)})|= \|h^{(1)}\|_0=m$. Since all non-zero entries of $h^{(1)}$  have magnitude larger than $\alpha/(t-1)$, we have
$$\alpha k \geq \|h^{(1)}\|_1 = \sum_{i\in \text{supp}(h^{(1)})} |h^{(1)}(i)| \geq \sum_{i\in \text{supp}(h^{(1)})} \alpha/(t-1) = m\alpha/(t-1). $$
Namely $m\leq k(t-1)$. Hence, \eqref{eq:main1} still holds. Besides, $\|h_{\max(k)} + h^{(1)}\|_0 = k+m \leq tk$, we have
\begin{equation}\label{eq:AhAh}
\langle A(h_{\max(k)}+h^{(1)}), Ah\rangle \leq \|A(h_{\max(k)} + h^{(1)})\|_2\|Ah\|_2 \leq \sqrt{1+\delta}\|h_{\max(k)} + h^{(1)}\|_2(\varepsilon+\eta).
\end{equation}
Again by \eqref{eq:main1}, we apply Lemma \ref{lm:mean} by setting $s=k(t-1)-m$, we can express $h^{(2)}$ as a weighted mean: $h^{(2)}=\sum_{i=1}^N\lambda_i u_i$, where $u_i$ is $(k(t-1)-m)$-sparse and \eqref{eq:after_lemma_u_i} still holds. Hence,
$$\|u_i\|_2\leq\sqrt{\|u_i\|_0}\|u_i\|_\infty \leq\sqrt{k(t-1)-m}\|u_i\|_\infty\leq\sqrt{k(t-1)}\|u_i\|_\infty \leq\sqrt{k/(t-1)}\alpha.$$
Now we suppose $1\geq \mu\geq 0, c\geq0$ are to be determined. Denote $\beta_i=h_{\max(k)}+h^{(1)}+\mu u_i$, then we still have \eqref{eq:mainth2}. Similarly to the proof of Theorem \ref{th:main}, since $h_{\max(k)}, h^{(1)}, u_i$ are $k$-, $m$-, $(k(t-1)-m)$-sparse vectors, respectively, we know $\beta_i = h_{\max(k)} + h^{(1)} + \mu u_i$, $\sum_{j=1}^N\lambda_j \beta_j-c\beta_i-\mu h = (1-\mu - c) (h_{\max(k)} + h^{(1)}) - c\mu u_i$ are all $tk$ sparse vectors.

Suppose $x=\|h_{\max(k)}+h^{(1)}\|_2$, $P=\frac{2\|\beta_{-\max(k)}\|_1}{\sqrt{k}}$, then
$$\quad \|u_i\|_2\leq \sqrt{k/(t-1)}\alpha\leq \frac{\|h_{\max(k)}\|_2}{\sqrt{(t-1)}}+\frac{2\|\beta_{-\max(k)}\|_1}{\sqrt{k(t-1)}}\leq\frac{\|h_{\max(k)}+h^{(1)}\|_2}{\sqrt{t-1}}+\frac{2\|\beta_{-\max(k)}\|_1}{\sqrt{k(t-1)}}=\frac{x+P}{\sqrt{t-1}}.$$
We still use the $\ell_2$ identity \eqref{eq:thmain_identity}. Set $c=1/2$, $\mu=\sqrt{t(t-1)}-(t-1)$ and take the difference of the left- and right-hand sides of \eqref{eq:thmain_identity}, we get
\begin{equation*}
\begin{split}
0  = 
  &\sum_{i=1}^N\lambda_i\left\|A\left((h_{\max(k)}+h^{(1)} + \mu h^{(2)})-\frac{1}{2}(h_{\max(k)} + h^{(1)} + \mu u_i)\right)\right\|_2^2 - \sum_{i=1}^N\frac{\lambda_i}{4}\|A\beta_i\|_2^2\\
  = &\sum_{i=1}^N\lambda_i\left\|A\left((\frac{1}{2}-\mu)(h_{\max(k)}+h^{(1)})-\frac{\mu}{2}u_i+\mu h\right)\right\|_2^2 - \sum_{i=1}^N\frac{\lambda_i}{4}\|A\beta_i\|_2^2\\
  = & \sum_{i=1}^N\lambda_i\left\|A\left((\frac{1}{2}-\mu)(h_{\max(k)}+h^{(1)})-\frac{\mu}{2}u_i\right)\right\|_2^2 + \mu^2\|Ah\|_2^2 \\
    & + 2\left\langle A\left((\frac{1}{2}-\mu)(h_{\max(k)}+h^{(1)}) - \frac{\mu}{2}h^{(2)}\right) , \mu Ah\right\rangle -\sum_{i=1}^N\frac{\lambda_i}{4}\|A\beta_i\|_2^2\\
  = & \sum_{i=1}^N\lambda_i\left\|A\left((\frac{1}{2}-\mu)(h_{\max(k)}+h^{(1)})-\frac{\mu}{2}u_i\right)\right\|_2^2  \\
    & + \mu(1-\mu)\left\langle A(h_{\max(k)}+h^{(1)}) , Ah\right\rangle -\sum_{i=1}^N\frac{\lambda_i}{4}\|A\beta_i\|_2^2.
\end{split}
\end{equation*}
Now since $\beta_i$, $(\frac{1}{2} - \mu)(h_{\max(k)} + h^{(1)}) - \frac{\mu}{2}u_i$ are all $tk$-sparse vectors, we apply the definition of $\delta_{tk}^A$ and also \eqref{eq:AhAh} to get
\begin{equation}\label{eq:longformula}
\begin{split}
0 \leq & (1+\delta) \sum_{i=1}^N\lambda_i\left((\frac{1}{2}-\mu)^2\|h_{\max(k)}+h^{(1)}\|_2^2+\frac{\mu^2}{4}\|u_i\|_2^2\right) + \mu(1-\mu)\sqrt{1+\delta}\|h_{\max(k)}+h^{(1)}\|_2(\varepsilon+\eta)\\
        & -(1-\delta)\sum_{i=1}^N\frac{\lambda_i}{4}\left(\|h_{\max(k)}+h^{(1)}\|_2^2+\mu^2\|u_i\|_2^2\right)\\
    = & \sum_{i=1}^N\lambda_i\left[\left((1+\delta)(\frac{1}{2}-\mu)^2-(1-\delta)\cdot\frac{1}{4}\right)\left\|h_{\max(k)}+h^{(1)}\right\|_2^2+\frac{1}{2}\delta\mu^2\|u_i\|_2^2\right] \\
    &+ \mu(1-\mu)\sqrt{1+\delta}\left\|h_{\max(k)}+h^{(1)}\right\|_2(\varepsilon+\eta)\\
   \leq & \left[(\mu^2-\mu)+\delta\left(\frac{1}{2}-\mu +(1+\frac{1}{2(t-1)})\mu^2\right)\right]x^2 + \left[\mu(1-\mu)\sqrt{1+\delta}(\varepsilon+\eta)+\frac{\delta\mu^2 P}{t-1}\right] x + \frac{\delta\mu^2 P^2}{2(t-1)}\\
    = & -t\left((2t-1)-2\sqrt{t(t-1)}\right)\left(\sqrt{\frac{t-1}{t}}-\delta\right)x^2+\left[\mu^2\sqrt{\frac{t}{t-1}}\cdot\sqrt{1+\delta}(\varepsilon+\eta) + \frac{\delta\mu^2P}{t-1}\right]x+\frac{\delta\mu^2P^2}{2(t-1)}\\
 = & \frac{\mu^2}{t-1}\left[-t\left(\sqrt{\frac{t-1}{t}}-\delta\right) x^2 +\left(\sqrt{t(t-1)(1+\delta)}(\varepsilon+\eta)+\delta P\right)x + \frac{\delta P^2}{2}\right],
\end{split}
\end{equation}
which is an second-order inequality for $x$. By solving this inequality we get
\begin{eqnarray*}
x&\leq& \frac{\left(\sqrt{t(t-1)(1+\delta)}(\varepsilon+\eta)+\delta P\right) + \sqrt{\left(\sqrt{t(t-1)(1+\delta)}(\varepsilon+\eta)+\delta P\right)^2+2t(\sqrt{(t-1)/t}-\delta)\delta P^2}}{2t(\sqrt{(t-1)/t}-\delta)}\\
 & \leq & \frac{\sqrt{t(t-1)(1+\delta)}}{t(\sqrt{(t-1)/t}-\delta)}(\varepsilon+\eta) + \frac{2\delta  + \sqrt{2t(\sqrt{(t-1)/t}-\delta)\delta}}{2t(\sqrt{(t-1)/t}-\delta)}P.
\end{eqnarray*}
Finally, note that $\|h_{-\max(k)}\|_1\leq \|h_{\max(k)}\|_1+P\sqrt{k}$, by Lemma 5.3 in \cite{Cai_Zhang}, we obtain $\|h_{-\max(k)}\|_2\leq \|h_{\max(k)}\|_2+P$, so
\begin{eqnarray*}
\|h\|_2 & = & \sqrt{\|h_{\max(k)}\|_2^2+\|h_{-\max(k)}\|_2^2}\\
 & \leq & \sqrt{\|h_{\max(k)}\|_2^2 + (\|h_{\max(k)}\|_2+P)^2}\\
 & \leq & \sqrt{2\|h_{\max(k)}\|_2^2} + P\\
 & \leq & \sqrt{2} x + P\\
 & \leq & \frac{\sqrt{2t(t-1)(1+\delta)}}{t(\sqrt{(t-1)/t}-\delta)}(\varepsilon+\eta) + \left(\frac{\sqrt{2}\delta  + \sqrt{t(\sqrt{(t-1)/t}-\delta)\delta}}{t(\sqrt{(t-1)/t}-\delta)}+1\right)\frac{2\|\beta_{-\max(k)}\|_1}{\sqrt{k}}\\
 & = & \frac{\sqrt{2(1+\delta)}}{1-\sqrt{t/(t-1)}\delta}(\varepsilon+\eta) + \left(\frac{\sqrt{2}\delta  + \sqrt{t(\sqrt{(t-1)/t}-\delta)\delta}}{t(\sqrt{(t-1)/t}-\delta)}+1\right)\frac{2\|\beta_{-\max(k)}\|_1}{\sqrt{k}},
\end{eqnarray*}
which finished the proof.

When $tk$ is not an integer, again we define $t' = \lceil tk \rceil/k$, then $t'>t$ and $\delta_{t'k}^A= \delta_{tk}^A < \sqrt{\frac{t-1}{t}} < \sqrt{\frac{t'-1}{t'}}$. We can prove the result by 
working on $\delta_{t'k}^A$. 

For the inequality on $\hat\beta^{DS}$ \eqref{eq:hatbetaDS}, the proof is similar. Define $h=\hat\beta^{DS} - \beta$. We have the following inequalities
$$\|A^TAh\|_\infty \leq \|A^T(A\hat\beta^{\ell_2} - y)\|_\infty + \|A^T(y - A\beta)\|_\infty\leq \eta+\varepsilon,$$
\begin{equation}\label{eq:AhAhDS}
\langle A(h_{\max(k)}+h^{(1)}), Ah\rangle = \langle h_{\max(k)}+h^{(1)}, A^TAh\rangle \leq \|h_{\max(k)}+h^{(1)}\|_1(\varepsilon+\eta) \leq \sqrt{tk}(\varepsilon+\eta)\|h_{\max(k)+h^{(1)}}\|_2,
\end{equation}
instead of \eqref{eq:Ah} and \eqref{eq:AhAh}. We can prove \eqref{eq:hatbetaDS} basically the same as the proof above except that we use \eqref{eq:AhAhDS} instead of \eqref{eq:AhAh} when we go from the third term to the fourth term in \eqref{eq:longformula}.
$\square$

\medskip\noindent
{\bf Proof of Proposition \ref{pr:gaussian}.}
By a small extension of Lemma 5.1 in \cite{Cai_l1}, we have $\|z\|_2\leq\sigma\sqrt{n+2\sqrt{n\log n}}$ with probability at least $1-1/n$; $\|A^Tz\|_\infty\leq \sigma\sqrt{2(1+\delta_1^A)\log p} \leq 2\sigma\sqrt{\log p}$ with probability at least $1-1/\sqrt{\pi \log p}$. Then the Proposition is immediately implied by Theorem \ref{th:noisy}. \quad $\square$

\medskip\noindent
{\bf Proof of Proposition \ref{pr:oracle_signal}.}
The proof of Proposition \eqref{pr:oracle_signal} is similar to that of Theorem 4.1 in \cite{Cai_Zhang} and Theorem 2.7 in \cite{Candes_Oracle}. 

First, as in the proof of Proposition \ref{pr:gaussian}, we have $\|A^Tz\|_\infty \leq \lambda/2$ with probability at least $1/\sqrt{\pi \log n}$. In the rest proof, we will prove \eqref{eq:oracle_inequality} in the event that $\|A^Tz\|_\infty\leq \lambda/2$. Define 
$$ K(\xi, \beta) = \gamma\|\xi\|_0 + \|A\beta - A\xi\|_2^2,\quad \gamma= \frac{\lambda^2}{8} = 2\sigma^2\log p.$$
Let $\bar \beta = \arg\min_\xi K(\xi, \beta)$. Since $K(\bar\beta, \beta) \leq K(\beta, \beta)$, we have $\gamma\|\bar\beta\|_0 \leq \gamma\|\beta\|_0$, which means $\bar{\beta}$ is $k$-sparse.

 Now we introduce the following lemma which can be regarded as an extension of Lemma 4.1 in \cite{Cai_Zhang}. 
\begin{Lemma}\label{lm:delta_k, delta_sk}
Suppose $A\in \mathbb{R}^{n\times p}$, $k\geq 2$ is an integer, $s>1$ is real and $sk$ is integer. Then we have $\delta_{sk}^A \leq (2s-1)\delta_k^A$. Similarly, suppose $\mathcal{M}:\mathbb{R}^{m\times n} \to \mathbb{R}^q$ is a linear map, $r \geq 2$ is an integer, $s>1$ is real and $sr$ is integer. Then we have $\delta_{sr}^{\mathcal{M}} \leq (2s-1)\delta_r^\mathcal{M}$.
\end{Lemma}
We omit the proof here as the proof of Lemma 4.1 in \cite{Cai_Zhang} can still apply to this lemma.

By Lemma \ref{lm:delta_k, delta_sk}, we can see when $1< t<2$, 
$$\delta_{2k}^A \leq (2\frac{2k}{\lceil tk\rceil} - 1)\delta_{\lceil tk\rceil}^A \leq (4/t - 1)\delta_{tk}^A \leq \sqrt{t/(t-1)}\delta_{tk}^A.$$
When $t\geq2$, $\delta_{2k}^A\leq \delta_{tk}^A$, which means \begin{equation}\label{eq:delta_2k, delta_tk}
\delta_{2k}^A \leq \sqrt{t/(t-1)}\delta_{tk}^A,
\end{equation}
whenever $t\geq 4/3$.

Next, we have 
$$\|\bar \beta - \beta\|_2^2 \leq \frac{1}{1-\delta_{2k}^A}\|A\bar\beta - A\beta\|_2^2 \leq \frac{1}{1-\sqrt{t/(t-1)}\delta_{tk}^A}\|A\bar\beta - A\beta\|_2^2.$$
With a small edition on Lemma 5.4 in \cite{Cai_Zhang} and Lemma 3.5 in \cite{Candes_Oracle}, we have
$$\|A^T(y-A\bar{\beta})\|_\infty \leq \|A^T(y- A\beta)\|_\infty + \|A^TA(\beta - \bar{\beta})\|_\infty\leq \lambda.$$
Since $\bar{\beta}$ is $k$-sparse, we can apply Theorem \ref{th:noisy} by plugging $\beta$ by $\bar{\beta}$ and get
$$\|\hat\beta - \bar\beta\|_2\leq \frac{\sqrt{2t\|\bar\beta\|_0}}{{1 - \sqrt{t/(t-1)}\delta_{tk}^A}}2\lambda. $$
Hence,
\begin{equation*}
\begin{split}
\|\hat{\beta} - \beta\|_2^2 & \leq 2\|\hat \beta - \bar \beta\|_2^2 + 2\|\bar \beta - \beta\|_2^2 \leq \frac{16t\|\bar \beta\|_0\lambda^2}{(1-\sqrt{t/(t-1)}\delta_{tk}^A)^2} + \frac{2}{1-\sqrt{t/(t-1)}\delta_{tk}^A} \|A\bar{\beta} - A\beta\|_2^2\\
& \leq \frac{128t}{(1-\sqrt{t/(t-1)}\delta_{tk}^A)^2}K(\bar{\beta}, \beta).
\end{split}
\end{equation*}
Suppose $\beta' = \sum_{i=1}^p\beta \cdot 1_{\{|\beta_i| > \mu\}}$, where $\mu = \sqrt{\frac{\gamma}{1+\delta_{k}^A}}$. Then
\begin{equation*}
\begin{split}
K(\bar\beta, \beta) & \leq K(\beta', \beta) \leq \gamma\sum_{i=1}^p 1_{\{|\beta_i|>\mu\}}+\|A\beta' - A\beta\|_2^2\\
& \leq \gamma \sum_{i=1}^p 1_{\{|\beta_i| > \mu\}} + (1+\delta_k^A)\sum_{i=1}^p1_{\{|\beta_i| \leq \mu\}}|\beta_i|^2 \leq \sum_{i=1}^p \min(\gamma, (1+\delta_k^A)|\beta_i|^2)\\
& \leq 2\log p\sum_{i=1}^p \min(\sigma^2, |\beta_i|^2).
\end{split}
\end{equation*}
Therefore, we have proved \eqref{eq:oracle_inequality} in the event that $\|A^Tz\|_\infty\leq \lambda/2$. \quad $\square$

\medskip\noindent
{\bf Proof of Theorem \ref{th:counterexample}.}
For any $\varepsilon>0$ and $k\geq 5/\varepsilon$, suppose $p\geq2tk$, $m'=((t-1)+\sqrt{t(t-1)})k$, $m$ is the largest integer strictly smaller than $m'$. Then $m< m'$ and $m'-m\leq1$. Since $t\geq 4/3$, we have $m'\geq k$. Define 
$$\beta_1 = \sqrt{k+\frac{mk^2}{m'^2}}^{-1}(\overbrace{1,\cdots, 1}^{k}, \overbrace{-\frac{k}{m'},\cdots, -\frac{k}{m'}}^{m},0,\cdots,0)\in\mathbb{R}^p,$$
then $\|\beta_1\|_2=1$. We define linear map $A:\mathbb{R}^p\to\mathbb{R}^p$, such that for all $\beta\in\mathbb{R}^{p}$,
$$ A\beta = \sqrt{1+\sqrt{\frac{t-1}{t}}}\left(\beta - \langle \beta_1, \beta\rangle \beta_1\right). $$
Now for all $\lceil tk \rceil$-sparse vector $\beta$, 
$$\|A\beta\|_2^2=\left(1+\sqrt{\frac{t-1}{t}}\right)(\beta-\langle\beta_1, \beta\rangle\beta_1)^T(\beta-\langle\beta_1, \beta\rangle\beta_1)=\left(1+\sqrt{\frac{t-1}{t}}\right)\left(\|\beta\|_2^2-|\langle\beta_1, \beta\rangle|^2\right).$$
Since $\beta$ is $\lceil tk\rceil $-sparse, by Cauchy-Schwarz Inequality, 
\begin{eqnarray*}
0\leq|\langle\beta_1, \beta\rangle|^2 &\leq& \|\beta\|_2^2\cdot\|\beta_1\cdot 1_{\text{supp}(\beta)}\|_2^2\\
& \leq& \|\beta\|_2^2\|\beta_{1, \max(\lceil tk\rceil )}\|_2^2 = \|\beta\|_2^2 \cdot \frac{m'^2+k(\lceil tk\rceil-k)}{m'^2+mk} \\
& \leq & \frac{m'^2 +k^2(t-1) + k}{m'^2+m'k}\cdot\frac{1}{1-\frac{k(m'-m)}{m'^2+m'k}}\|\beta\|_2^2\\
& = & \frac{m'^2 +k^2(t-1)}{m'^2+m'k}\cdot\frac{m'^2 +k^2(t-1) + k}{m'^2 +k^2(t-1)}\cdot\frac{1}{1-\frac{k(m'-m)}{m'^2+m'k}}\|\beta\|_2^2\\
& = & 2\sqrt{t-1}(\sqrt{t}-\sqrt{t-1})\cdot (1+\frac{1}{tk})\cdot\frac{1}{1-\frac{1}{2k}}\|\beta\|_2^2\\
& \leq & \left(2\sqrt{t(t-1)} - 2(t-1)\right) \cdot (1+\frac{5}{2k})\|\beta\|_2^2\\
& \leq & \left(2\sqrt{t(t-1)}-2(t-1)+\frac{5}{2k}\right)\|\beta\|_2^2.
\end{eqnarray*}
We used the fact that $m'\geq k$, $0<m'-m\leq1$ and
\begin{equation*}
\begin{split}
\frac{m'^2 +k^2(t-1)}{m'^2+m'k} & =  \frac{\left((t-1) + \sqrt{t(t-1)}\right)^2 + t -1}{\left((t-1) + \sqrt{t(t-1)}\right)^2 + \left((t-1) + \sqrt{t(t-1)}\right)}\\
& = \frac{(t-1) \left(t-1 + t  +2\sqrt{t(t-1)} + 1\right)}{\sqrt{t(t-1)}\left(\sqrt{t} + \sqrt{(t-1)}\right)^2}\\
& = \frac{2\sqrt{t-1}}{\sqrt{t} + \sqrt{t-1}} = 2\sqrt{t-1}\left(\sqrt{t} - \sqrt{t-1}\right)
\end{split}
\end{equation*}
 above. Hence, 
$$ \left(1+\sqrt{\frac{t-1}{t}}\right)\|\beta\|_2^2 \geq \|A\beta\|_2^2\geq \left(1-\sqrt{\frac{t-1}{t}} - \left(1+\sqrt{\frac{t-1}{t}}\right)\frac{5}{2k}\right)\|\beta\|_2^2\geq\left(1-\sqrt{\frac{t-1}{t}}-\varepsilon\right)\|\beta\|_2^2,$$
which implies $\delta_{tk}^A \leq \sqrt{(t-1)/t} +\varepsilon$.

Now we consider 
$$\beta_0 = (\overbrace{1,\cdots, 1}^k, 0,\cdots,0)\in\mathbb{R}^p,$$
$$\gamma_0 = (\overbrace{0,\cdots,0}^k, \overbrace{\frac{k}{m'}, \cdots, \frac{k}{m'}}^{m},0,\cdots, 0).$$
Note that $A\beta_1=0$, so $A\beta_0 = A\gamma_0$. Besides, $\beta_0$ is $k$-sparse and $\|\gamma_0\|_1<\|\beta_0\|_1$.
\begin{itemize}
\item In the noiseless case, i.e. $y=A\beta_0$, the $\ell_1$ minimization method \eqref{eq:signalmini} fails to exactly recover $\beta_0$ through $y$ since $y=A\gamma_0$, but $\|\gamma_0\|_1<\|\beta_0\|_1$.

\item In the noisy case, i.e. $y=A\beta_0+z$, assume that $\ell_1$ minimization method \eqref{eq:signalmini} can stably recover $\beta_0$ with constraint $\mathcal{B}_z$. Suppose $\hat\beta_z$ is the solution of $\ell_1$ minimization, then $\lim_{z\to 0} \hat\beta_z=\beta_0$. Note that $y - A(\hat\beta_z-\beta_0+\gamma_0) = y-A\hat\beta_z\in\mathcal{B}_z$, by the definition of $\hat\beta_z$, we have $\|\hat\beta_z-\beta_0+\gamma_0\|_1\geq \|\hat\beta_z\|_1$. Let $z\to 0$, it contradicts that $\|\gamma_0\|_1<\|\beta_0\|_1$. Therefore, $\ell_1$ minimization method \eqref{eq:signalmini} fails to stably recover $\beta_0$.
\quad $\square$
\end{itemize}

\medskip\noindent
{\bf  Proof of Proposition \ref{pr:t<1}. } We use the technical tools developed in Cai and Zhang \cite{Cai_Zhang2} to prove this result. We begin by introducing another important concept in the RIP framework - restricted orthogonal constants (ROC) proposed in \cite{Candes_Decoding}.
\begin{Definition}
Suppose $A\in\mathbb{R}^{n\times p}$, define the restricted orthogonal constants (ROC) of order $k_1, k_2$ as the smallest non-negative number $\theta_{k_1, k_2}^A$ such that 
$$|\langle A\beta_1, A\beta_2\rangle | \leq \theta_{k_1, k_2}^A\|\beta_1\|_2\|\beta_2\|_2,$$
for all $k_1$-sparse vector $\beta_1\in\mathbb{R}^p$ and $k_2$-sparse vector $\beta_2\in\mathbb{R}^p$ with disjoint supports.
\end{Definition}
Based on Theorem 2.5 in \cite{Cai_Zhang2}, 
\begin{equation}\label{eq:ROCcondition}
\delta_{tk}^A+\frac{2k-tk}{tk}\theta_{tk, tk}^A<1
\end{equation}
is a sufficient condition for exact recovery of all $k$-sparse vectors. By Lemma 3.1 in \cite{Cai_Zhang2}, $\theta_{tk,tk}^A\leq 2\delta_{tk}^A$ when $tk$ is even; $\theta_{tk, tk}^A\leq \frac{2tk}{\sqrt{(tk)^2 -1}}\delta_{tk}^A$ when $tk$ is odd. Hence,
$$\delta_{tk}^A+\frac{2k-tk}{tk}\theta_{tk, tk}^A\leq \frac{4-t}{t}\delta_{tk}^A,\quad \text{ when $tk$ is even;}$$
$$\delta_{tk}^A+\frac{2k-tk}{tk}\theta_{tk,tk}^A\leq\left(1+\frac{4k-2tk}{\sqrt{(tk)^2 - 1}}\right)\delta_{tk}^A,\quad \text{ when $tk$ is odd.}$$
The proposition is implied by the inequalities above and \eqref{eq:ROCcondition}. \quad $\square$

\medskip\noindent
{\bf  Proof of Proposition \ref{pr:t<4/3}.} The idea of the proof is quite similar to Theorem 3.2 by Cai and Zhang \cite{Cai_Zhang}. Define 
$$\gamma = \frac{1}{\sqrt{2k}}(\overbrace{1, \cdots, 1}^{2k}, 0, \cdots,0),$$
\begin{equation*}
\begin{split}
A:\mathbb{R}^p &\to \mathbb{R}^p\\
 \beta &\mapsto \frac{2}{\sqrt{4-t}}\left(\beta - \langle \beta,\gamma\rangle \gamma\right).
\end{split}
\end{equation*}
Now for all non-zero $\lceil tk\rceil$-sparse vector $\beta\in\mathbb{R}^p$, 
$$\|A\beta\|_2^2 = \frac{4}{4-t}\langle \beta-\langle \beta,\gamma\rangle\gamma, \beta-\langle \beta,\gamma\rangle\gamma\rangle=\frac{4}{4-t}(\|\beta\|_2^2 - \langle \beta,\gamma\rangle^2).$$
We can immediately see $\|A\beta\|_2^2 \leq (1+t/(4-t))\|\beta\|_2^2$. On the other hand by Cauchy-Schwarz's inequality,
$$\langle \beta, \gamma\rangle^2 = \langle \beta, \gamma\cdot 1_{\{supp(\beta)\}} \rangle^2 \leq \|\beta\|_2^2(\sum_{i\in supp(\beta)} \gamma_i^2)\leq \|\beta\|_2^2\cdot \frac{\lceil tk\rceil}{2k}.$$
For $k>1/\varepsilon$, we have 
$$ \|A\beta\|_2^2 \geq \frac{4}{4-t}(1-\frac{\lceil tk\rceil}{2k})\|\beta\|_2^2 \geq \frac{4}{4-t}(1-\frac{tk}{2k}-\varepsilon/2)\|\beta\|_2^2 > (1-\frac{t}{4-t} - \varepsilon)\|\beta\|_2^2.$$
Therefore, we must have $\delta_{tk}^A = \delta_{\lceil tk\rceil}^A < t/(4-t) + \varepsilon$.

Finally, we define
$$\beta_0 = (\overbrace{1, \cdots, 1}^k, 0,\cdots,0),\quad \beta_0'=(\overbrace{0,\cdots, 0}^k, \overbrace{-1,\cdots, -1}^k, 0,\cdots, 0).$$
Then $\beta_0, \beta_0'$ are both $k$-sparse, and $y=A\beta_0=A\beta_0'$. There's no way to recover both $\beta_0, \beta_0'$ only from $(y, A)$.
$\square$

\end{document}